\documentclass[useAMS,usenatbib]{mn2e}
\usepackage{psfig}
\usepackage[a4paper]{hyperref}
\usepackage{amssymb}
\usepackage{aas_macros}

\newcommand{\rmsub}[2]{#1_{\rm #2}} 

\title[Transiting exoplanets WASP-1b and WASP-2b]{WASP-1b and WASP-2b: Two new transiting exoplanets detected with SuperWASP and SOPHIE}
\author[A. Collier Cameron et al]
{
A. Collier Cameron$^{1}$\thanks{E-mail:acc4@st-and.ac.uk},
 F. Bouchy$^{12,13}$,
G. H\'ebrard$^{12}$,
P. Maxted$^{5}$,
D. Pollacco$^{2}$,
\newauthor
F. Pont $^{10}$,
I. Skillen$^{8}$,
B. Smalley$^{5}$,
R. A. Street$^{2}$,
R.G. West $^{3}$,
D.M. Wilson $^{5}$,
\newauthor
S. Aigrain$^{6}$,
D.J. Christian$^{2}$,
W.I. Clarkson$^{4,15}$,
B. Enoch$^{4}$,
A. Evans$^{5}$,
\newauthor
A. Fitzsimmons$^{2}$,
M. Fleenor$^{16}$,
M. Gillon$^{14}$,
C.A. Haswell$^{4}$,
L. Hebb$^{1}$,
C. Hellier$^{5}$,
\newauthor
S.T. Hodgkin$^{6}$,
K. Horne$^{1}$,
J. Irwin$^{6}$,
S.R. Kane$^{7}$,
F.P. Keenan$^{2}$,
B. Loeillet$^{11}$,
\newauthor
T.A. Lister$^{1,5}$,
M. Mayor$^{10}$,
C. Moutou$^{11}$,
A.J. Norton$^{4}$,
J. Osborne$^{3}$,
N. Parley$^{4}$,
\newauthor
D. Queloz$^{10}$,
R. Ryans$^{2}$,
A.H.M.J. Triaud$^{1}$,
S. Udry$^{10}$,
and
P.J. Wheatley $^{9}$
\\
\\
$^{1}$School of Physics and Astronomy, University of St Andrews, North Haugh, St Andrews, Fife KY16 9SS, UK.\\
$^{2}$ARC, Main Physics Building, School of Mathematics \&\ Physics, Queen's University, University Road, Belfast, BT7 1NN, UK.\\
$^{3}$Department of Physics and Astronomy, University of Leicester, Leicester, LE1 7RH, UK.\\
$^{4}$Department of Physics and Astronomy, The Open University, Milton Keynes, MK7 6AA, UK.\\
$^{5}$Astrophysics Group, Keele University, Staffordshire, ST5 5BG.\\
$^{6}$Institute of Astronomy, University of Cambridge, Madingley Road, Cambridge, CB3 0HA, UK.\\
$^{7}$Department of Astronomy, University of Florida, 211 Bryant Space Science Center, Gainesville, FL 32611-2055, USA.\\
$^{8}$Isaac Newton Group of Telescopes, Apartado de Correos 321, E-38700 Santa Cruz de la Palma, Tenerife, Spain. \\
$^{9}$Department of Physics, University of Warwick, Coventry CV4 7AL, UK.\\
$^{10}$Observatoire de Gen\`eve, Universit\'e de Gen\`eve, 51 Ch. des Maillettes, 1290 Sauverny, Switzerland.\\
$^{11}$Laboratoire d'Astrophysique de Marseille, BP 8, 13376 Marseille Cedex 12, France\\
$^{12}$Institut d'Astrophysique de Paris, CNRS (UMR 7095) --  Universit\'e Pierre \&\ Marie Curie, 98$^{bis}$ bvd. Arago, 75014 Paris, France\\
$^{13}$Observatoire de Haute-Provence, 04870 St Michel l'Observatoire, France\\
$^{14}$Institut d'Astrophysique et de G\'eophysique,  Universit\'e
 de Li\`ege,  All\'ee du 6 Ao\^ut 17,  4000 Li\`ege, Belgium\\
$^{15}$STScI, 3700 San Martin Drive, Baltimore, MD 21218, USA\\
$^{16}$Volunteer Observatory, 10305 Mantooth Lane Knoxville, TN 37932, USA
}
\begin{document}

\date{Accepted 2006 November 28. Received 2006 November 7; in original form 2006 September 22}

\pagerange{\pageref{firstpage}--\pageref{lastpage}} \pubyear{2006}

\maketitle

\label{firstpage}

\begin{abstract}
We have detected low-amplitude radial-velocity variations in two stars, 
USNO-B1.0 1219-0005465
(GSC 02265-00107 = WASP-1) and USNO-B1.0 0964-0543604 (GSC 00522-01199 = WASP-2). Both stars were identified as being likely host stars of transiting exoplanets in the 2004 SuperWASP wide-field transit survey. Using the newly-commissioned radial-velocity spectrograph SOPHIE at the Observatoire de Haute-Provence, we found that both objects exhibit reflex orbital radial-velocity variations with amplitudes characteristic of planetary-mass companions and in-phase with the photometric orbits. Line-bisector studies rule out faint blended binaries as the cause of either the radial-velocity variations or the transits. We perform preliminary spectral analyses of the host stars, which together with their radial-velocity variations and fits to the transit light curves, yield estimates of the planetary masses and radii. WASP-1b and WASP-2b have orbital periods of 2.52 and 2.15 days respectively. Given mass estimates for their F7V and K1V primaries we derive planet masses 0.80 to 0.98  and 0.81 to 0.95 times that of Jupiter respectively. WASP-1b appears to have an inflated radius of at least 1.33~$\rmsub{R}{Jup}$, whereas WASP-2b has a radius in the range 0.65 to 1.26~$\rmsub{R}{Jup}$. 
\end{abstract}


\section{Introduction}

Extra-solar planets that transit their parent stars are of key interest because their masses and radii can be determined directly, providing clues to their internal compositions \citep{guillot2006}. They define the mass-radius-separation relation for irradiated giant planets \citep{mazeh2005}. They provide unique insights into their thermal properties (e.g. \citealt{charbonneau2005}; \citealt{deming2005}; \citealt{deming2006}) and the chemical compositions of their atmospheres (\citealt{charbonneau2002}; \citealt{vidalmadjar2003}; \citealt{vidalmadjar2004}). 

The first exoplanet found to exhibit transits, HD 209458b (\citealt{charbonneau2000}; \citealt{henry2000}), was initially discovered using the radial-velocity method that has to date yielded the vast majority of the 210 known exoplanets. Its inflated radius \citep{brown2001} presents a challenge to theories of the structure and evolution of irradiated exoplanets. New radial-velocity discoveries are routinely subjected to careful photometric followup at around the predicted times of transit. Indeed, recent radial-velocity surveys targeting bright stars of high metallicity with the specific goal of discovering new hot Jupiters  have revealed two new transiting planets in the last two years (\citealt{sato2005}; \citealt{bouchy2005}). Transiting planets thus comprise roughly 10\% of all the hot Jupiters with orbital periods under 5 days or so, which is consistent with expectations for randomly-oriented orbits. 

The complementary approach is to look for transits first, then to seek evidence of their
planetary nature in radial-velocity followup observations. The OGLE project pioneered this approach, which has to date yielded five transiting exoplanet candidates for which
radial-velocity variations have been detected with amplitudes indicative of planetary-mass companions (\citealt{konacki2003}; \citealt{bouchy2004}, \citealt{konacki2004}; \citealt{pont2004}; \citealt{konacki2005}). 

Several teams have embarked on ultra-wide field searches for transiting exoplanets. In this complementary approach, small-aperture CCD imaging systems coupled to commercial camera optics secure light curves of millions of stars. The TrES survey
has so far yielded two transiting planets (\citealt{alonso2004}; \citealt{odonovan2006b}). The XO survey and HAT surveys have so far found one each (\citealt{mccullough2006}; \citealt{bakos2006}). Here we present the first two transiting exoplanets from the 
SuperWASP survey. The hardware, data analysis pipeline and archive methodology for this project are described in detail by  \citet{pollacco2006}.  

In a series of recent papers, \citet{christian2006}, \citet{street2006} and \citet{lister2006} presented transit candidates from the inaugural 2004 May--September observing season of the five SuperWASP wide-field survey cameras that were operating at that time. The light curves of some $1.1\times 10^6$ stars in the magnitude range $8<V<13$ were searched for periodic shallow transits, and several dozen plausible candidates were identified for detailed followup. The candidate selection methodology is described in detail in these papers and by \citet{cameron2006}.


\section{SOPHIE and OFXB Observations}

\begin{table}
\caption[]{Target list for SOPHIE radial-velocity followup programme. The 1SWASP identifiers give the J2000 stellar coordinates. The second column gives the number $\rmsub{N}{cl}$ of SOPHIE observations that were required to classify and eliminate each non-planetary system, and the total number of observations for planet-bearing systems.
Half the astrophysical false positives were eliminated after a single observation. Five more were eliminated after a second observation, leaving 8 narrow, single-lined targets.}
\label{tab:targlist}
\begin{center}
\begin{tabular}{lrl}
\hline\\	

1SWASP ID	& $\rmsub{N}{cl}$ & SOPHIE followup classification	\\

\hline\\	

J161732.90+242119.0	& 3 &	No detectable RV variation	\\
J165949.13+265346.1	& 1 &	Rapid rotator	\\
J174118.30+383656.3	& 2 &	SB1	\\
J181252.03+461851.6	& 2 &	No detectable RV variation	\\
J183104.01+323942.7	& 1 &	SB2	\\
J184303.62+462656.4	& 1 & 	Late-type giant	\\
J203054.12+062546.4	& 9 &	Planet host: WASP-2	\\
J204125.28+163911.8	& 1 &	Late-type giant	\\
J205027.33+064022.9	& 3 &	No detectable RV variation	\\
J205308.03+192152.7	& 1 &	SB2	\\
J210318.01+080117.8	& 4 &	Line bisector variable	\\
J211608.42+163220.3	& 1 &	SB2	\\
J214151.03+260158.5	& 1 &	Late-type giant	\\
J215802.14+253006.1	& 1 &	SB2	\\
J222317.60+130125.8	& 1 &	Late-type giant	\\
J223320.44+370139.1	& 1 &	SB2	\\
J223651.20+221000.8	& 2 &	SB2	\\
J234318.41+295556.5	& 2 &	SB1	\\
J002040.07+315923.7	& 7 &	Planet host: WASP-1	\\
J005225.90+203451.2	& 3 &	No detectable RV variation	\\
J010151.11+314254.7	& 1 &	Rapid rotator	\\
J025500.31+281134.0	& 4 &	No detectable RV variation	\\
J031103.19+211141.4	& 1 &	Rapid rotator	\\
J051221.34+300634.9	& 2 &	Multiple system \\

\hline\\	
		
\end{tabular}
\end{center}
\end{table}

\begin{table*}
\caption[]{Journal of radial-velocity measurements of WASP-1 and WASP-2. The 1SWASP identifiers give the J2000 stellar coordinates of the photometric apertures; the USNO-B1.0 number denotes the star for which the radial-velocity measurements were secured. The uncertainties given here include 10.0 m s$^{-1}$ systematic error in added in quadrature to the formal photon-noise error. The fifth and sixth columns give the FWHM of the CCF dip and the contrast of the dip as a fraction of the weighted mean continuum level. The signal-to-noise ratio near 550 nm is given in column 6, and the spectral type of the cross-correlation mask in column 7. }
\label{tab:journal}
\begin{center}
\begin{tabular}{ccccccccl}

HJD & $\rmsub{t}{exp}$ &$\rmsub{V}{r}$ & FWHM & Contrast & S:N & Mask & Notes \\
       &           (s)          &  km s$^{-1}$    & km s$^{-1}$ & \% &        & Sp. type & \\
\hline\\
\multicolumn{8}{l}{\bf 1SWASP J002040.07+315923.7: USNO-B1.0 1219-0005465 = GSC 02265-00107 = WASP-1 } \\

2453979.6311 & 2400 & $-13.425 \pm 0.011$ & 9.7 & 25.3 & 37.7 & G2 & \\
2453980.5558 & 2400 & $-13.484 \pm 0.012$ & 9.7 & 21.7 & 36.5 & G2 & During transit\\
2453981.5649 & 2400 & $-13.587 \pm 0.016$ & 9.8 & 17.1 & 25.0 & G2 & Variable cloud \\
2453981.6752 & 1219 & $-13.520 \pm 0.021$ & 9.8 & 13.8 & 21.4 & G2 & Variable cloud \\
2453982.4167 & 2100 & $-13.376 \pm 0.014$ & 9.8 & 17.8 & 31.5 & G2 & \\
2453982.5843 & 2100 & $-13.412 \pm 0.013$ & 10.0 & 19.3 & 33.3 & G2 & \\
2453982.6758 & 2407 & $-13.398 \pm 0.012$ & 10.0 & 22.0 & 35.0 & G2 & \\
\\
\multicolumn{8}{l}{\bf 1SWASP J203054.12+062546.4: USNO-B1.0 0964-0543604 = GSC 00522-01199 = WASP-2} \\
2453981.5065 & 1906 & $-28.125 \pm 0.186$ & 8.9 & 2.2 & 12.4 & K5 & Variable to heavy cloud \\
2453982.3786 & 2500 & $-27.711 \pm 0.012$ & 93 & 19.1 & 37.3 & K5 & \\
2453982.4962 & 2500 & $-27.736 \pm 0.013$ & 9.0 & 17.0 & 39.4 & K5 & \\
2453991.3817 & 1200 & $-27.780 \pm 0.011$ & 8.8 & 30.2 & 45.0 & K5 & \\
2453991.5102 & 1200 & $-27.812 \pm 0.011$ & 8.8 & 29.8 & 41.6 & K5 & During transit \\
2453996.3529 & 1200 & $-28.037 \pm 0.011$ & 9.0 & 25.0 & 37.5 & K5 &  \\
2453996.4301 & 1200 & $-28.020 \pm 0.012$ & 9.0 & 24.4 & 35.4 & K5 &  \\
2453997.3824 & 1500 & $-27.723 \pm 0.012$ & 9.0 & 24.5 & 36.7 & K5 & \\
2453998.3415 & 1200 & $-27.987 \pm 0.011$ & 9.0 & 25.3 & 40.2 & K5 & \\
\hline\\
\end{tabular}
\end{center}
\label{default}
\end{table*}

We conducted a radial-velocity survey of a sample of high-priority SuperWASP transit candidates, using the newly-commissioned SOPHIE spectrograph \citep{bouchy2006sophie} on the 1.93-m telescope at the Observatoire de Haute-Provence during the four nights from 2006 August 31 to 2006 September 3, and between September 12 and 19 during the science verification phase of SOPHIE. SOPHIE is a bench-mounted, fibre-fed spectrograph built on the same design principles as the HARPS instrument \citep{pepe2004} on the ESO 3.6-m telescope at La Silla. The spectrograph's thermal environment is carefully controlled, with the aim of achieving radial-velocity measurements with stability better than 2 m s$^{-1}$. For the targets studied here, a radial-velocity precision of 10 to 15 m s$^{-1}$ is adequate to establish or reject the planetary nature of a transit candidate. We therefore elected to use SOPHIE's High-Efficiency (HE) mode, which has resolving power $\lambda/\Delta\lambda=35000$. The CCD detector records 39 spectral orders spanning the wavelength range from 387 to 694 nm. Radial velocities are determined by cross-correlation with a mask spectrum matched to the spectral type of the target (\citealt{baranne1996}; \citealt{pepe2002}). Automatic data reduction at the telescope allows highly efficient candidate selection and data assessment.

The target list was drawn from the candidate papers of  \citet{christian2006}, \citet{street2006} and \citet{lister2006}, and from further candidate lists covering different regions of the sky for which candidate papers are currently in preparation.  Table \ref{tab:targlist} gives a brief summary of all targets observed and the outcomes of the SOPHIE observations. 

A full investigation of the natures of the other objects observed during this radial-velocity study is currently in progress. The results of this investigation and its implications for improving our pre-selection criteria will be published in a companion paper. Here we present the two targets in the survey sample that have single, narrow-lined cross-correlation functions (CCF) with radial velocity variations in agreement with an oscillation precisely phased with the ephemeris predicted from SuperWASP data. The amplitudes of these radial velocity oscillations are less than a few hundred m s$^{-1}$ and no significant line-bisector variations are detected (see Section~\ref{sec:blends}). Thus we can conclude that each of these two stars harbours a transiting planet. The radial-velocity measurements for the two stars are listed in Table 1.


A complete transit of WASP-2 was observed using a CCD camera with $R$-band filters on the 60-cm telescope of the Observatoire Fran\c{c}ois-Xavier Bagnoud at St-Luc (OFXB), on the night of 2006 September 12/13 UT. A full transit of WASP-1 was observed on the morning of 2006 October 2 UT using an SBIG ST10XME CCD camera  with $R$-band filter on the 0.35-m Schmidt-Cassegrain Telescope at the Volunteer Observatory at Knoxville, Tennessee.

\section{Stellar parameters}
\label{sect:spectra}

The extracted SOPHIE spectra were used for a preliminary analysis using the {\sc uclsyn} spectral synthesis package and ATLAS9 models without convective
overshooting (Castelli et al. 1997). The H$\alpha$, Na{\sc i} $D$ and Mg{\sc i} $b$ lines were used as diagnostics of both $\rmsub{T}{eff}$ and $\log g$. The abundances do not appear to be substantially different from solar. We used these values to infer the radii and masses of the stars, as listed in Table~\ref{tab:uclsyn}.  For WASP-1, comparison with the stellar evolution models of \citet{girardi2000} gives maximum-likelihood values $M_*=1.15\rmsub{M}{\odot}$ and $R_*=1.24\rmsub{R}{\odot}$, but many models with $1.06<M_*/\rmsub{M}{\odot}<1.39$ and $1.04<R_*/\rmsub{R}{\odot}<1.92$ satisfy the spectroscopic constraints, because the main sequence is very wide in that temperature range.  The radius and mass estimates for WASP-2 are better constrained. We have also used the available BV and 2MASS photometry to estimate $\rmsub{T}{eff}$ using the Infrared Flux Method \citep{blackwell1977}, which gave results in agreement with that obtained from the spectral analysis.

\section{Planetary parameters}

The formal precision of the radial-velocity observations depends on the signal-to-noise ratio of the spectrum, the sharpness and density of the stellar lines, and the scattered-light background in the instrument. The formal errors on the velocity measures are given by the semi-empirical estimator $\rmsub{\sigma}{RV}=1.7\sqrt{\rm FWHM}$/(S:N*Contrast). The Contrast parameter quantifies the contrast of the CCF peak against the additional background signal from light leakage into the as-yet incomplete spectrograph enclosure. Because we did not use simultaneous thorium-argon wavelength calibration, the accuracy of the radial-velocity measurements is limited by the stability of the spectrograph over the 2 to 3 hours between successive thorium-argon calibration exposures, considering that its thermal control was not yet optimized. Tests performed during the commissioning of the instrument indicate that the velocity drift during a night is typically in the range $\pm 10$ m s$^{-1}$. Taking into consideration additional uncertainties coming from the wavelength solution of the H.E. mode and the guiding noise, we estimated that during our run, the systematic RV errors were 10 m/s. Although the wavelength calibration may vary only slowly during a given night, the drifts may reasonably be expected to be uncorrelated from one night to the next. Guiding noise should be uncorrelated even between successive observations. Since most targets were observed only once per night, we treat the additional systematic error as uncorrelated. We therefore used the quadrature sum of the formal and systematic errors (Table~\ref{tab:journal}) for all model fits.

\begin{table}
\caption[]{Stellar parameters derived from preliminary spectral analyses with {\sc uclsyn}. Masses and radii are derived from the Padua models of \citet{girardi2000} assuming [Fe/H]$=0.1\pm 0.2$ to reflect the uncertainty in the metallicity.}
\label{tab:uclsyn}
\begin{center}
\begin{tabular}{lcc}
\hline\\
Parameter & WASP-1 & WASP-2 \\
\hline\\
GSC & 02265-00107 & 00522-01199 \\
 WASP $V$ (mag) & 11.79 & 11.98 \\
 Spectral type       & F7V & K1V \\
 $\rmsub{T}{eff} $ (K) & $6200\pm 200$ & $5200\pm 200$ \\
 $\log g$               & $4.3 \pm 0.3$ & $4.3 \pm 0.3$ \\
 $M_V$ (mag)              & $3.9\pm 0.4$ & $6.2\pm 0.5$ \\ 
 $M_*/M_\odot$ & $1.15^{+0.24}_{-0.09}$ & $0.79^{+0.15}_{-0.04}$ \\
 $R_*/R_\odot$ & $1.24^{+0.68}_{-0.20}$ & $0.78\pm 0.06$ \\
 \hline\\
\end{tabular}
\end{center}
\end{table}

\begin{table}
\caption[]{Results of simultaneous minimum-$\chi^2$ circular-orbit fits to the photometric and radial-velocity data for WASP-1 and WASP-2. The parameters of the lightcurve
model are given in terms of the radius of the star and planet ($R_*$ and $\rmsub{R}{p}$, respectively), the separation of the stars ($a$) and the inclination ($i$).  The total number of degrees of freedom, including the $\rmsub{N}{RV}$ radial-velocity measurements, is $\rmsub{N}{df}$. The contribution of the radial-velocity data to the value of $\chi^2$ is $\chi^2_{\rm RV}$, using the errors given in Table~\ref{tab:journal}. Data in transit are given reduced weight. Standard errors on the parameters are derived using a
bootstrap analysis as described in the text.}
\label{tab:params}
\begin{tabular}{lcc}
\hline\\
Parameter                        & WASP-1b & WASP-2b \\
                                 & & \\
\hline\\
Transit HJD        & $2453912.514 \pm 0.001$ & $2453991.5146 \pm 0.0044 $ \\
Period (days)      & $2.51995 \pm 0.00001$ &  $2.152226 \pm 0.000004$ \\
$\gamma$ (km s$^{-1}$)    &$-13.503 \pm 0.009$& $-27.863 \pm 0.007$ \\
$K_1$ (m s$^{-1}$)        &$115 \pm 11$        & $155 \pm 7 $ \\
$a$ (AU) & 0.0369\,--\,0.0395 & 0.0296\,--\,0.0318 \\
$b = a\,\cos i/R_*$ &0\,--\,0.8         & 0\,--\,0.8 \\
$\rmsub{R}{p}/R_*$    &0.093\,--\,0.104   & 0.119\,--\,0.140 \\
$R_*/a$             &0.168\,--\,0.260   & 0.086\,--\,0.132 \\
$\rmsub{N}{df}$                  & 961               & 1013 \\
$\chi^2$                  & 1420\,--\,1449  & 1627.2\,--\,1647.1 \\
$\rmsub{N}{RV}$                  &7                  & 9 \\
$\chi^2_{\rm RV}$   & 11.6              & 13.4 \\
\noalign{\smallskip}
$M_* (M_{\odot})$    & (1.06\,--\,1.39) & (0.73\,--\,0.94)  \\
$\rmsub{M}{p}/\rmsub{M}{Jup}$ & (0.80\,--\,0.98)$\pm 0.11$ &(0.81 \,--\,0.95)$\pm 0.04$  \\
$\rmsub{R}{p}/\rmsub{R}{Jup}$ & 1.33\,--\,2.53 & 0.65\,--\,1.26 \\
\hline\\
\end{tabular}
\end{table}

The transits of WASP-1 and WASP-2 can be timed with a precision of about
20 minutes from the 2004 SuperWASP data set. The first and last transits
in this data set are separated by about 120 days, so the accumulated
uncertainties in the transit timings at the epoch of the 2006 SOPHIE, OFXB and Volunteer Observatory observations are no more than a few hours. There is no ambiguity in number of cycles between the 2004 and 2006 data sets. The OFXB and Volunteer Observatory transit observations establish the improved photometric ephemerides in Table~\ref{tab:params}.

\begin{figure*}
\begin{center}
\begin{tabular}{ll}
\psfig{figure=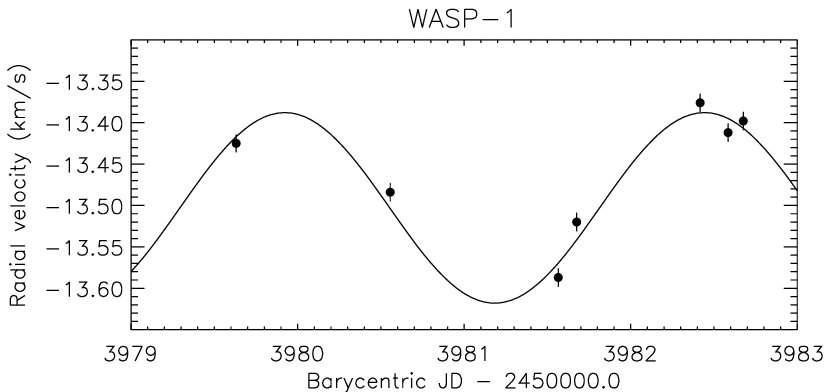,width=8.5cm} &
\psfig{figure=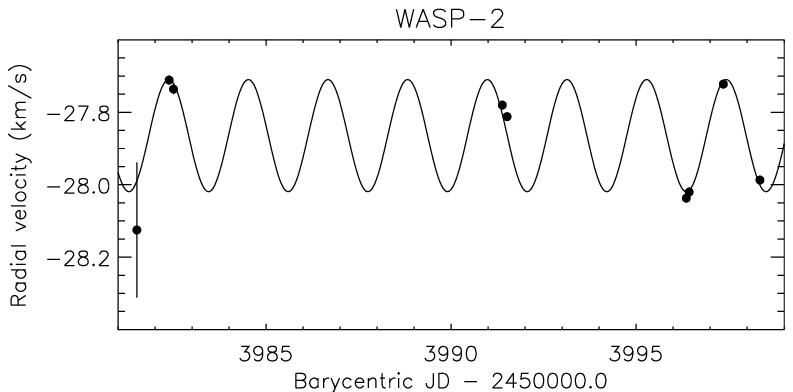,width=8.5cm} \\
\psfig{figure=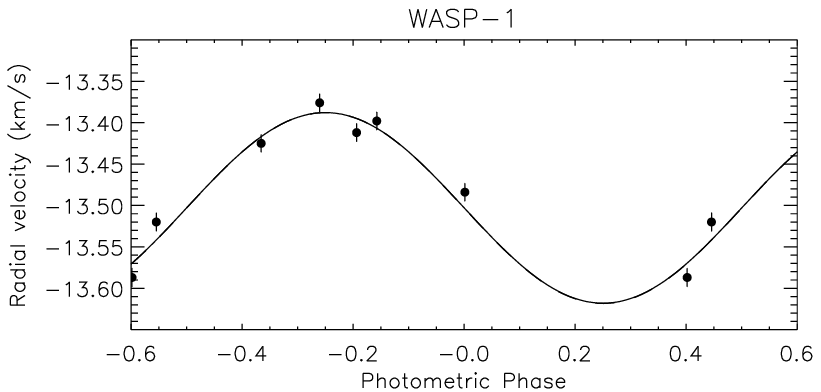,width=8.5cm} &
\psfig{figure=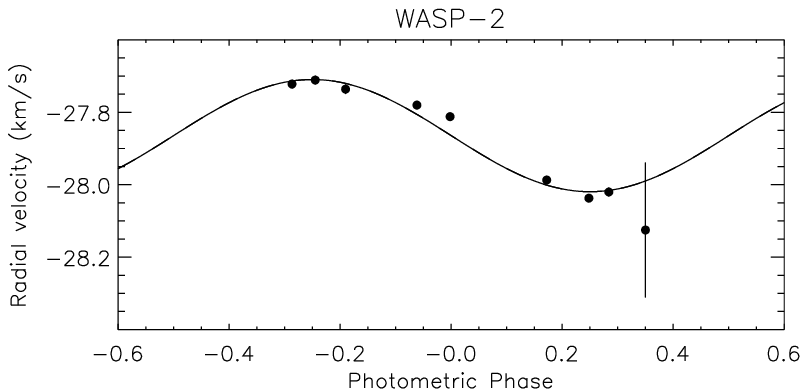,width=8.5cm}
\end{tabular}
\caption[]{Radial velocities of WASP-1and WASP-2 plotted against barycentric Julian date (upper) and folded on the photometric ephemerides (lower). In all panels the solid line represents the best-fitting circular-orbit solution.}
\label{fig:rvphase}
\end{center}
\end{figure*}

We estimated the stellar and planetary radii and the planetary masses by minimising $\chi^2$ for the photometric and radial-velocity measurements simultaneously with respect to  the analytic model of \citet{mandel2002} for small planets ($\rmsub{R}{p}/\rmsub{R}{*}<0.1$), assuming that the planets have circular orbits. In modelling the SuperWASP photometry we used linear limb darkening coefficients $u = 0.51$ and 0.63 \citep{vanhamme1993} for WASP-1 and WASP-2 respectively. For the more precise data from OFXB and Volunteer Observatory we used the 4-coefficient model of \citet{claret2000}. 

We used a bootstrap analysis to estimate the errors on the parameters of the model. This entails taking the residuals from the optimum light-curve fit, 
applying an arbitrary phase shift and restoring the model transit at phase zero. 
This preserves both outliers and the correlated  noise characteristics of the WASP data.
The synthetic radial-velocity data are generated by sampling the best fit radial-velocity
curve at the observed phases and adding gaussian random deviates with the 
appropriate standard error. These synthetic data are then fitted repeatedly, to recover the distributions of the fitted parameter values. 

There is a strong degeneracy between the impact parameter $b=a\cos i/R_*$ and the parameter $R_*/a$ when fitting planet transit lightcurves of the quality presented here. We therefore present the results of minimum-$\chi^2$ fits with the value of $b$ fixed at values $b=0$ (corresponding to the minimum radius of the star) and $b=0.8$ for both stars.  Models with $b>0.8$ give noticeably worse fits to the lightcurves.

\begin{figure}
\begin{center}
\psfig{figure=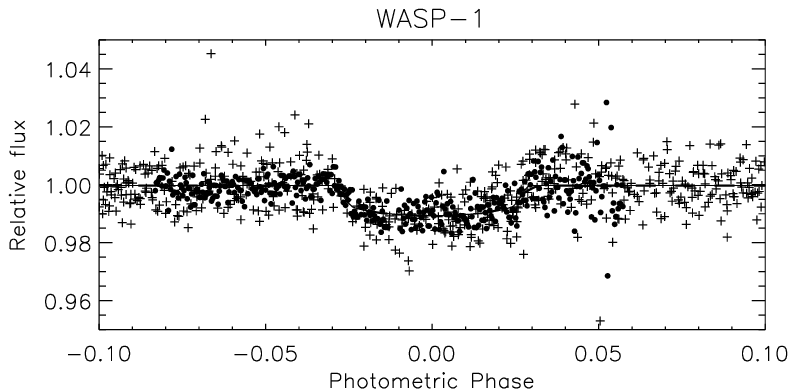,width=8.5cm}
\psfig{figure=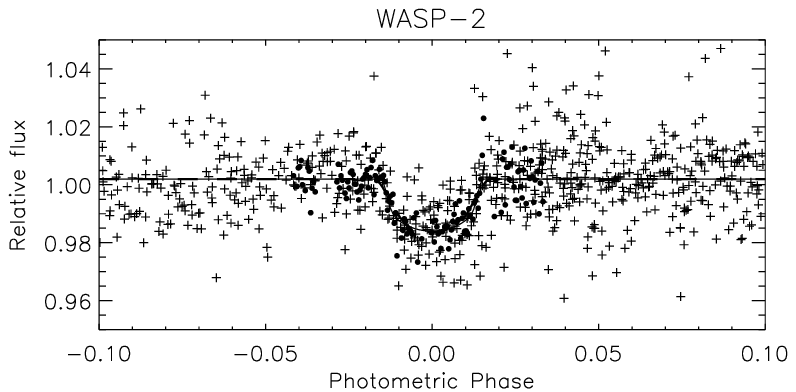,width=8.5cm}
\caption[]{Transit profiles of WASP-1 (upper) and WASP-2 (lower), fitted with the 
optimum limb-darkened models with parameters as in Table~\ref{tab:params}. Small
symbols denote 2004 season SuperWASP photometry; filled circles show the Volunteer Observatory light curve of 2006 October 1 (WASP-1) and the OFXB light curve of 2006 September 12 (WASP-2).}
\label{fig:folded}
\end{center}
\end{figure}

The radial-velocity measurements and minimum-$\chi^2$ fits are shown as a
function of orbital phase in Fig.~\ref{fig:rvphase}. The sinusoidal variation in radial
velocity is clearly seen. The light curves and minimum-$\chi^2$ fits to the phases
around the transits are shown in Fig.~\ref{fig:folded}. The parameters of the best fits are
given in Table~\ref{tab:params}. An additional 50 m s$^{-1}$ was added in quadrature to the uncertainties of radial-velocity observations during transits, to allow for the Rossiter-McLaughlin effect (\citealt{rossiter1924}; \citealt{mclaughlin1924}). 

In spite of this, the contribution of the radial-velocity data to the total value of $\chi^2$ in both stars is larger than expected. WASP-1 yields a spectroscopic $\chi^2_s=11.6$ for 5 degrees of freedom, while WASP-2 gives  $\chi^2_s=13.4$ for 7 degrees of freedom. 
More extensive observations will be required to establish the cause of this small
additional ``jitter'' in the radial-velocity measurements. A common cause of such radial-velocity ``jitter" is rotationally-modulated distortion of the line profiles arising from magnetic activity in the stellar chromosphere and photosphere.  While the S:N in the bluest orders of the SOPHIE spectra is insufficient for us to measure the chromospheric emission cores in the Ca II HK lines in any meaningful way, both stars have narrow CCFs that suggest $v\sin i$ values less than $\sim 5$ km s$^{-1}$ and hence low to moderate activity levels. We allow for this additional variability in the bootstrap error analysis by adding 12 m s$^{-1}$ of additional radial-velocity jitter to the synthetic radial-velocity data. We note that this amount of jitter is within the ranges determined empirically for F and K stars with low to moderate levels of chromospheric activity \citep{wright2005}, and conclude that chromospheric activity is a likely cause.

Also given in Table~\ref{tab:params} are the masses and radii of WASP-1b and WASP-2b derived using the best-fit model parameters.  Despite the ambiguities in the light-curve solution it is clear that the data
presented show that WASP-1 and WASP-2 have planetary-mass companions with gas-giant radii.

\section{Faint blended-binary scenarios}
\label{sec:blends}

Although many types of stellar binary system can mimic a planet-like 
transit signal, most are easily eliminated. Strong rotational broadening
implies tidal synchronisation by a massive companion. Grazing double or single-lined stellar binaries reveal their nature after one or two observations.
Triple systems can, however, mimic transiting planets in a way that is  difficult to eliminate (\citealt{torres2004}; \citealt{odonovan2006a}). A bright single F, G or K star with a much fainter, physically associated eclipsing-binary companion will apparently produce shallow eclipses. If the bright primary has narrow absorption lines, the faint binary's lines will produce apparent periodic velocity shifts in the combined spectrum, in phase with the photometric orbit. The apparent shift occurs because the broadened lines of the tidally-synchronised primary of  the eclipsing binary are Doppler shifted by orbital motions into the wings of the composite line profile. 

Several tests can reveal this type of system. Because the spectral types of the bright star and the cooler eclipsing binary are very different, cross-correlation masks of different spectral types tend to yield different orbital velocity amplitudes \citep{santos2002}. We computed the  CCF with different masks without significant change in the radial-velocity values. 

Because the apparent Doppler shift arises through a variable asymmetry in the line wings, line-bisector analysis of the cross-correlation function is also an effective probe for this type of system \citep{queloz2001}. When the fainter, moving spectrum is broadened by rotation, the induced velocity shift is sufficiently weak that the apparent velocity variation tends to be small unless the binary's lines are strong enough to give a clearly-visible asymmetry in the combined line profile. The change in line-bisector velocity from the wings to the core of the line tends to be greater than the induced velocity amplitude. We measured the asymmetries of the cross-correlation function peaks using the line-bisector method of \citet{queloz2001}. For WASP-1 and WASP-2 we found the scatter in bisector velocities from the wings to the core of the CCF profile to be substantially less than the measured orbital velocity amplitude, and uncorrelated with orbital phase. A third candidate, 1SWASP J210318.01+080117.8, failed this test and was eliminated as a planet candidate.

Therefore, we can eliminate the scenario ``bright single star plus faint, late-type short-period eclipsing binary'' with confidence for WASP-1 and WASP-2. While other more intricate scenarios are possible, we were not able to contrive any that could explain both the photometric and the velocity signals while remaining credible.

At the 380 and 140 pc distances of WASP-1 and WASP-2, it should be possible to resolve such binary companions at separations of a few tens of AU or more with the help of adaptive optics. As an additional line of defence against this type of astrophysical false positive, we secured high-resolution $H$-band images of both targets with the NAOMI adaptive-optics system on the 4.2-m William Herschel Telescope on the nights of 2006 September 6 and 7. Images with corrected FWHM = 0.25 arcsec reveal that WASP-1 has a {\bf stellar} companion 4.7 arcsec to the north and 3.7 magnitudes fainter at $H$. The SOPHIE fibre aperture has a diameter of 3 arcsec. Seeing and guiding errors may allow some of the companion's light into the fibre, but we can be fairly confident that the contamination is not significant at visible wavelengths. Even if the companion were an eclipsing binary, it would be unlikely to mimic the radial-velocity signature of a planet, and indeed the line-bisector analysis eliminates this possibility. NAOMI H-band images taken on 2006 September 7  with corrected FWHM=0.2 arcsec, and using the OSCA coronagraph system, show that WASP-2 has a stellar companion 2.7 magnitudes fainter at $H$ located 0.7" to the east. This falls within the SOPHIE fibre aperture. Additional NAOMI images secured during transit on 2006 September 10 20:00 to 20:20 UT with 0.2 arcsec corrected FWHM showed no sign of the $\sim$1.5-mag deep eclipse in the companion that would be needed to mimic a transit. Future AO observations should reveal whether these faint stellar companions (which are the only objects visible besides WASP-1 and WASP-2 in their respective NAOMI fields of view) are chance alignments or common-proper-motion companions.

\begin{figure}
\begin{center}
\psfig{figure=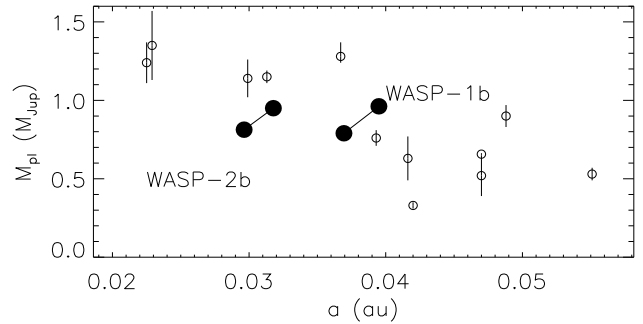,width=8.5cm}
\caption[]{Mass-orbital separation diagram for the 14 known transiting exoplanets. WASP-1b and WASP-2b follow the general trend with only high-mass planets surviving at small separations. (From {\it  http://obswww.unige.ch/$\sim$pont/TRANSITS.htm})}
\label{fig:sepmass}
\end{center}
\end{figure}

\section{Discussion and conclusions}

We  have detected the presence of radial-velocity variations in two exoplanetary transit candidates. Our preliminary analysis yields masses between 0.8 and 1.0 $\rmsub{M}{Jup}$ for both planets. WASP-1b and WASP-2b lie between the ``hot Jupiters" and the ``very hot Jupiters" in the mass-orbital separation diagram for transiting hot Jupiters,   (Fig.~\ref{fig:sepmass}).Their intermediate masses appear consistent with the general trend toward high masses at the smallest orbital separations noted by \citet{mazeh2005}. WASP-2b in particular lies close to the minimum separation at which planets in this mass range survive, making it  a good candidate for future mass-loss studies. The radius of WASP-1b, which orbits an F7V star, is poorly constrained by the SuperWASP data alone, but appears to be at least 1.33~$\rmsub{R}{Jup}$. WASP-1b seems thus to be an expanded, low-density planet, similar to HD 209458b and HAT-P-1b \citep{bakos2006}. Additional photometry of WASP-2b, which orbits a K1 dwarf at a slightly smaller orbital separation yields a radius close to that of Jupiter, suggesting a substantially higher density. Additional high-precision photometry is needed to refine the radii and densities of both planets; indeed, precise radius estimates have recently been derived from high-precision photometry by \citet{shporer2006} and \citet{charbonneau2006}, confirming the oversized nature of WASP-1b.

\section*{Acknowledgments}

We are grateful to all the staff of Observatoire de Haute Provence for their efforts, their efficiency and their support on the new instrument SOPHIE and for the photometric observation with the 1.20-m telescope, and in particular H. Le Coroller and R. Giraud. We acknowledge the award of Director's Discretionary Time at Haute-Provence for additional observations of WASP-2 in SOPHIE science verification time, without which the detection could not have been secured. We extend our special thanks to the team of the Observatoire Fran\c{c}ois-Xavier Bagnoud at St-Luc, and in particular to Brice-Olivier Demory and Fr{\'e}d{\'e}ric Malmann. Additional partial observations of the 2006 September 12  transit of WASP-2 were secured at short notice with the OHP 1.2-m telescope. The WASP project is funded and operated by Queen's University Belfast, the Universities of Keele, St Andrews and Leicester, the Open University, the Isaac Newton Group, the Instituto de Astrofisica de Canarias, the South African Astronomical Observatory and by PPARC. In recognition of the considerable regional support given to the WASP project on La Palma, we would like to associate the pseudonym Garafia-1 with the planet of WASP-1.  AMHJT was supported by an Undergraduate Research Bursary from the the Nuffield Foundation. This publication makes use of data products from the Two Micron All Sky Survey, which is a joint project of the University of Massachusetts and the Infrared Processing and Analysis Center/California Institute of Technology, funded by the National Aeronautics and Space Administration and the National Science Foundation. This research has made use of  the VizieR catalogue access tool, CDS, Strasbourg, France. We are grateful to the referee, Dr. Chris Tinney, for numerous helpful suggestions to improve and clarify the manuscript. 

\bibliographystyle{mn2e}

\begin{thebibliography}{}

\bibitem[\protect\citeauthoryear{Alonso et al.}{2004}]{alonso2004} 
Alonso R., et al., 2004, ApJ, 613, L153 

\bibitem[\protect\citeauthoryear{{Bakos}, {Noyes}, {Kovacs}, {Latham},
  {Sasselov}, {Torres}, {Fischer}, {Stefanik}, {Sato}, {Johnson}, {Pal},
  {Marcy}, {Butler}, {Esquerdo}, {Stanek}, {Lazar}, {Papp}, {Sari} \&
  {Sipocz}}{{Bakos} et~al.}{2006}]{bakos2006}
{Bakos} G.~A.,  et al.,
  2006, \apj, In  press (arXiv: astro-ph/0609369)


\bibitem[\protect\citeauthoryear{{Baranne}, {Queloz}, {Mayor}, {Adrianzyk},
  {Knispel}, {Kohler}, {Lacroix}, {Meunier}, {Rimbaud} \& {Vin}}{{Baranne}
  et~al.}{1996}]{baranne1996}
{Baranne} A., et al.,
  1996, \aaps, 119, 373

\bibitem[\protect\citeauthoryear{{Blackwell} \& {Shallis}}{{Blackwell} \&
{Shallis}}{1977}]{blackwell1977}
{Blackwell} D.~E.,  {Shallis} M.~J.,  1977, \mnras, 180, 177

\bibitem[\protect\citeauthoryear{Bouchy et al.}{2004}]{bouchy2004} 
Bouchy F., Pont F., Santos N.~C., Melo C., Mayor M., Queloz D., Udry S., 
2004, A\&A, 421, L13 

\bibitem[\protect\citeauthoryear{Bouchy et al.}{2005}]{bouchy2005} 
Bouchy F., et al., 2005, \aap, 444, L15 

\bibitem[\protect\citeauthoryear{{Bouchy} et~al.}{{Bouchy} et~al.}{2006}]{bouchy2006sophie}
{Bouchy} F.,  {The Sophie Team} 2006, in {Arnold} L.,  {Bouchy} F.,   {Moutou}
  C.,  eds, Tenth Anniversary of 51 Peg-b: Status of and prospects for hot
  Jupiter studies,
pp 319--325

\bibitem[\protect\citeauthoryear{Brown et al.}{2001}]{brown2001} 
Brown T.~M., Charbonneau D., Gilliland R.~L., Noyes R.~W., Burrows A., 
2001, ApJ, 552, 699 

\bibitem[\protect\citeauthoryear{{Castelli}, {Gratton} \& {Kurucz}}{{Castelli} et~al.}{1997}]{castelli1997}
{Castelli} F., {Gratton} R.~G., {Kurucz} R.~L., 1997, \aap, 318, 841

\bibitem[\protect\citeauthoryear{Charbonneau et 
al.}{2000}]{charbonneau2000} Charbonneau D., Brown T.~M., Latham D.~W., 
Mayor M., 2000, ApJ, 529, L45 

\bibitem[\protect\citeauthoryear{{Charbonneau}, {Brown}, {Noyes} \&
  {Gilliland}}{{Charbonneau} et~al.}{2002}]{charbonneau2002}
{Charbonneau} D.,  {Brown} T.~M.,  {Noyes} R.~W.,    {Gilliland} R.~L.,  2002,
  \apj, 568, 377

\bibitem[\protect\citeauthoryear{{Charbonneau}, {Allen}, {Megeath}, {Torres},
  {Alonso}, {Brown}, {Gilliland}, {Latham}, {Mandushev}, {O'Donovan} \&
  {Sozzetti}}{{Charbonneau} et~al.}{2005}]{charbonneau2005}
{Charbonneau} D.,  et al.,  2005, \apj, 626, 523

\bibitem[\protect\citeauthoryear{Charbonneau et 
al.}{2006}]{charbonneau2006} Charbonneau D., Winn J.~N., Everett M.~E., 
Latham D.~W., Holman M.~J., Esquerdo G.~A., O'Donovan F.~T., 2006, \apj, Submitted 
(arXiv:astro-ph/0610589)

\bibitem[\protect\citeauthoryear{Christian et 
al.}{2006}]{christian2006} Christian D.~J., et al., 2006, \mnras, 372, 
1117 

\bibitem[\protect\citeauthoryear{{Claret}}{{Claret}}{2000}]{claret2000}
{Claret} A.,  2000, \aap, 363, 1081

\bibitem[\protect\citeauthoryear{{Collier Cameron}, {Pollacco}, {Street},
  {Lister}, {West}, {Wilson}, {Pont}, {Clarkson}, {Christian}, {Enoch},
  {Evans}, {Fitzsimmons}, {Haswell}, {Hellier}, {Hodgkin}, {Horne}, {Irwin},
  {Kane}, {Keenan}, {Norton}, {et al.}}{{Collier Cameron} et~al.}{2006}]{cameron2006}
{Collier Cameron} A.,  et al.,  2006, \mnras, In press.  (arXiv: astro-ph/0609418)

\bibitem[\protect\citeauthoryear{{Deming}, {Harrington}, {Seager} \&
  {Richardson}}{{Deming} et~al.}{2006}]{deming2006}
{Deming} D.,  {Harrington} J.,  {Seager} S.,    {Richardson} L.~J.,  2006,
  \apj, 644, 560

\bibitem[\protect\citeauthoryear{{Deming}, {Seager}, {Richardson} \&
  {Harrington}}{{Deming} et~al.}{2005}]{deming2005}
{Deming} D.,  {Seager} S.,  {Richardson} L.~J.,    {Harrington} J.,  2005,
  \nat, 434, 740

\bibitem[\protect\citeauthoryear{{Girardi}, {Bressan}, {Bertelli} \&
  {Chiosi}}{{Girardi} et~al.}{2000}]{girardi2000}
{Girardi} L.,  {Bressan} A.,  {Bertelli} G.,    {Chiosi} C.,  2000, \aaps, 141,
  371

\bibitem[\protect\citeauthoryear{{Guillot}, {Santos}, {Pont}, {Iro}, {Melo} \&
  {Ribas}}{{Guillot} et~al.}{2006}]{guillot2006}
{Guillot} T.,  et al.,  2006, \aap, 453, L21

\bibitem[\protect\citeauthoryear{Henry et al.}{2000}]{henry2000} 
Henry G.~W., Marcy G.~W., Butler R.~P., Vogt S.~S., 2000, ApJ, 529, L41 

\bibitem[\protect\citeauthoryear{Konacki et al.}{2003}]{konacki2003} Konacki M., Torres G., Jha S., Sasselov D.~D., 2003, \nat, 421, 507 

\bibitem[\protect\citeauthoryear{Konacki et 
al.}{2004}]{konacki2004} Konacki M., et al., 2004, ApJ, 609, L37 

\bibitem[\protect\citeauthoryear{Konacki et 
al.}{2005}]{konacki2005} Konacki M., Torres G., Sasselov D.~D., Jha 
S., 2005, ApJ, 624, 372 

\bibitem[\protect\citeauthoryear{{Lister}, {West}, {Wilson}, {Collier Cameron},
  {Clarkson}, {Street}, {Enoch}, {Parley}, {Christian}, {Kane}, {Evans},
  {Fitzsimmons}, {Haswell}, {Hellier}, {Hodgkin}, {Horne}, {Irwin}, {Keenan},
  {Norton}, {Osborne}, {et al.}}{{Lister} et~al.}{2006}]{lister2006}
{Lister} T.~A., et al.,  2006, \mnras, Submitted

\bibitem[\protect\citeauthoryear{{Mandel} \& {Agol}}{{Mandel} \&
  {Agol}}{2002}]{mandel2002}
{Mandel} K.,  {Agol} E.,  2002, \apjl, 580, L171

\bibitem[\protect\citeauthoryear{{Mazeh}, {Zucker} \& {Pont}}{{Mazeh}
  et~al.}{2005}]{mazeh2005}
{Mazeh} T.,  {Zucker} S.,    {Pont} F.,  2005, \mnras, 356, 955

\bibitem[\protect\citeauthoryear{{McCullough}, {Stys}, {Valenti},
  {Johns-Krull}, {Janes}, {Heasley}, {Bye}, {Dodd}, {Fleming}, {Pinnick},
  {Bissinger}, {Gary}, {Howell} \& {Vanmunster}}{{McCullough}
  et~al.}{2006}]{mccullough2006}
{McCullough} P.~R.,  et al.,  2006, \apj, 648, 1228

\bibitem[\protect\citeauthoryear{{McLaughlin}}{{McLaughlin}}{1924}]{mclaughlin%
1924}
{McLaughlin} D.~B.,  1924, \apj, 60, 22

\bibitem[\protect\citeauthoryear{O'Donovan et 
al.}{2006b}]{odonovan2006b} O'Donovan F.~T., et al., 2006b, ApJ, 651, 
L61 

\bibitem[\protect\citeauthoryear{O'Donovan et 
al.}{2006a}]{odonovan2006a} O'Donovan F.~T., et al., 2006a, ApJ, 644, 
1237 

\bibitem[\protect\citeauthoryear{{Pepe}, {Mayor}, {Galland}, {Naef}, {Queloz},
  {Santos}, {Udry} \& {Burnet}}{{Pepe} et~al.}{2002}]{pepe2002}
{Pepe} F.,  et al.,  2002, \aap, 388, 632

\bibitem[\protect\citeauthoryear{{Pepe}, {Mayor}, {Queloz}, {Benz}, {Bonfils},
  {Bouchy}, {Curto}, {Lovis}, {M{\'e}gevand}, {Moutou}, {Naef}, {Rupprecht},
  {Santos}, {Sivan}, {Sosnowska} \& {Udry}}{{Pepe} et~al.}{2004}]{pepe2004}
{Pepe} F.,  {Mayor} M.,  {Queloz} D.,  et al.,  2004, \aap, 423, 385

\bibitem[\protect\citeauthoryear{Pollacco et 
al.}{2006}]{pollacco2006} Pollacco D.~L., et al., 2006, \pasp, 118, 
1407

\bibitem[\protect\citeauthoryear{Pont et al.}{2004}]{pont2004} 
Pont F., Bouchy F., Queloz D., Santos N.~C., Melo C., Mayor M., Udry S., 
2004, A\&A, 426, L15 

\bibitem[\protect\citeauthoryear{{Queloz}, {Henry}, {Sivan}, {Baliunas},
  {Beuzit}, {Donahue}, {Mayor}, {Naef}, {Perrier} \& {Udry}}{{Queloz}
  et~al.}{2001}]{queloz2001}
{Queloz} D., et al.,  2001, \aap, 379, 279

\bibitem[\protect\citeauthoryear{{Rossiter}}{{Rossiter}}{1924}]{rossiter1924}
{Rossiter} R.~A.,  1924, \apj, 60, 15

\bibitem[\protect\citeauthoryear{{Santos}, {Mayor}, {Naef}, {Pepe}, {Queloz},
  {Udry}, {Burnet}, {Clausen}, {Helt}, {Olsen} \& {Pritchard}}{{Santos}
  et~al.}{2002}]{santos2002}
{Santos} N.~C.,  et al.,  2002, \aap, 392, 215

\bibitem[\protect\citeauthoryear{{Sato}, {Fischer}, {Henry}, {Laughlin},
  {Butler}, {Marcy}, {Vogt}, {Bodenheimer}, {Ida}, {Toyota}, {Wolf}, {Valenti},
  {Boyd}, {Johnson}, {Wright}, {Ammons}, {Robinson}, {Strader},
  {et al.}}{{Sato} et~al.}{2005}]{sato2005}
{Sato} B.,  et al.,  2005, \apj, 633, 465

\bibitem[\protect\citeauthoryear{Shporer et 
al.}{2006}]{shporer2006} Shporer A., Tamuz O., Zucker S., Mazeh T., 
2006, \mnras, Submitted 
(arXiv:astro-ph/0610556 )

\bibitem[\protect\citeauthoryear{{Street}, {Christian}, {Clarkson}, {Collier
  Cameron}, {Enoch}, {Kane}, {Lister}, {West}, {Wilson}, {Evans},
  {Fitzsimmons}, {Haswell}, {Hellier}, {Hodgkin}, {Horne}, {Irwin}, {Keenan},
  {Norton}, {Osborne}, {Pollacco}, {et al.}}{{Street}  et~al.}{2006}]{street2006}
{Street} R.~A.,  et al.,  2006, \mnras, Submitted

\bibitem[\protect\citeauthoryear{Torres et al.}{2004}]{torres2004} 
Torres G., Konacki M., Sasselov D.~D., Jha S., 2004, ApJ, 614, 979 

\bibitem[\protect\citeauthoryear{{van Hamme}}{{van Hamme}}{1993}]{vanhamme1993}
{van Hamme} W.,  1993, \aj, 106, 2096

\bibitem[\protect\citeauthoryear{{Vidal-Madjar}, {Lecavelier des Etangs},
  {D{\'e}sert}, {Ballester}, {Ferlet}, {H{\'e}brard} \& {Mayor}}{{Vidal-Madjar}
  et~al.}{2003}]{vidalmadjar2003}
{Vidal-Madjar} A.,  et al.,  2003,
  \nat, 422, 143
  
\bibitem[\protect\citeauthoryear{{Vidal-Madjar}, {D{\'e}sert}, {Lecavelier des
  Etangs}, {H{\'e}brard}, {Ballester}, {Ehrenreich}, {Ferlet}, {McConnell},
  {Mayor} \& {Parkinson}}{{Vidal-Madjar} et~al.}{2004}]{vidalmadjar2004}
{Vidal-Madjar} A.,  et al.,  2004, \apjl, 604, L69

\bibitem[\protect\citeauthoryear{Wilson et al.}{2006}]{wilson2006} 
Wilson D.~M., et al., 2006, PASP, 118, 1245

\bibitem[\protect\citeauthoryear{Wright}{2005}]{wright2005} Wright 
J.~T., 2005, \pasp, 117, 657 

\end{thebibliography}

\bsp

\label{lastpage}

\end{document}